\documentclass[fleqn,11pt]{wlscirep}
\usepackage[utf8]{inputenc}
\usepackage[T1]{fontenc}
\usepackage{graphicx}
\usepackage{epstopdf}
\usepackage{dcolumn}
\usepackage{bm}
\usepackage{braket}
\usepackage{color}
\usepackage{xcolor}
\usepackage{ragged2e}
\usepackage{amsmath, bm}
\usepackage{amsmath}
\usepackage{amssymb}%
\usepackage{url} %
\usepackage{tcolorbox}
\usepackage{hyperref}
\usepackage{bookmark}
\usepackage{physics}

\graphicspath{{Figure/}}

\title{Pulse-by-pulse programmable synthesis of ultrafast optical waveforms}
\author[1,2,4,*]{Shilong Liu }
\author[4]{Gabriel Demontigny}
\author[4]{\'{E}mile Dessureault}
\author[4]{St\'{e}phane Virally}
\author[3,4,+]{Denis V. Seletskiy}

\affil[1] {State Key Laboratory of Precision Spectroscopy, and Hainan Institute, East China Normal University, Shanghai, 200062, China}
\affil[2] {Chongqing Key Laboratory of Precision Optics, Chongqing Institute of East China Normal University, Chongqing, 401120, China}
\affil[3]{Department of Physics and Astronomy, FemtoQ Laboratory, 1 University of New Mexico, MSC07 4220, Albuquerque, New Mexico 87131-0001, United States}
\affil[4]{femtoQ Lab, Department of Engineering Physics, Polytechnique Montr\'{e}al, Montr\'{e}al, Qu\'{e}bec H3T 1J4, Canada}
\affil[*]{corresponding dr.shilongliu@gmail.com}

\affil[+]{corresponding  denis.seletskiy@polymtl.ca}

\begin{abstract}
Programmable control of individual pulses in a high-repetition-rate (typically MHz) ultrafast pulse train is a long-standing goal for optical arbitrary waveform synthesis. Here, we report a programmable pulse-by-pulse shaper that enables deterministic spectral-temporal control of ultrafast pulses at a repetition rate of $\sim$ 20~MHz. By synchronizing an FPGA-driven electro-optic modulation on the stretched waveform in a temporal 4$f$ shaping system, the regime writes pulse-index-dependent spectral phase profiles onto individual pulses. We demonstrate three levels of programmable ability: zero-order phase coding that maps pulse-by-pulse phase sequences into double-slit-like spectral-temporal interference; first-order phase programming that produces arbitrary temporal trajectories with deterministic delay; and fractional-order phase engineering that generates programmable temporal breathing. By launching the shaped pulse train into a nonlinear fiber stage, the programmed temporal breathing is converted into one spectral breathing. We further construct a phase-level-dependent regime map of nonlinear spectral breathing, revealing transitions from weak single-envelope breathing to multi-peak spectral splitting and strongly breathing merged-spectrum dynamics, in agreement with numerical simulations. This pulse-by-pulse spectral-temporal synthesis platform establishes pulse index as a programmable degree of freedom for ultrafast pulse shaping and provides a route toward real-time and pulse resolved optical arbitary waveform synthesis.
\end{abstract}

\begin{document}
\flushbottom
\maketitle
\thispagestyle{empty}
\noindent

\section*{Introduction}

The ability to shape ultrafast optical fields has become a central tool in modern ultrafast photonics. By tailoring the spectral amplitude and phase of an ultrashort pulse, one can synthesize user-defined temporal waveforms and control light--matter interactions on femtosecond-to-picosecond time scales~\cite{weiner2011ultrafast}. This capability has enabled coherent control of quantum and molecular dynamics~\cite{warren1993coherent,brixner2004quantum}, optical arbitrary waveform generation~\cite{weiner2000femtosecond,cundiff2010optical}, temporal-mode engineering~\cite{ansari2018tailoring,davis2018measuring}, supercontinuum and few-cycle pulse shaping~\cite{dudley2006supercontinuum,demmler2011generation}, and photonic- radio frequency signal processing~\cite{torres2011space,yao2011photonic}. In these applications, pulse shaping provides a programmable interface for synthesizing optical fields in both the temporal and spectral domain, such as generations of the pulse varying from multiple cycles to few- cycle \cite{weiner2011ultrafast,iegorov2016direct,liu2025engineering}. 

Most established pulse-shaping platforms, however, are designed to apply one imprinted mask to many successive pulses. For example, conventional Fourier-transform pulse shapers based on spatial light modulators (SLMs) or LCOS devices provide spectral amplitude and phase control through thousands of independently addressable pixels, but their refresh rates are typically far below the repetition rates of pulse trains from the mode-locked laser~\cite{weiner2011ultrafast}. Faster approaches based on digital micromirror devices (DMDs)~\cite{gu2015digital}, acousto-optic modulators (AOMs)~\cite{dugan1997high,touil2022acousto}, and electro-optic modulation (EOM)~\cite{thomas2010fiber,yao2011photonic} can increase the update speed, yet they often operate in restricted control operations to predefined temporal waveform, such as amplitude-only modulation, binary, or burst-level waveform shaping \cite{weiner2011ultrafast,stummer2020programmable}. Time-domain electro-optic pulse shapers have demonstrated high-speed reconfigurable picosecond waveform synthesis, including binary phase-only filtering and multi-level phase-only temporal shaping~\cite{thomas2010fiber,huh2016fiber}. Nevertheless, it remains challenging to combine and show the ability in both writing and readout of the programmable ultrafast optical waveforms in the manner of pulse-by-pulse. This challenge is closely related to a long-standing goal in optical arbitrary waveform generation: to realize fully programmable optical waveforms that can be updated at the rate of individual pulses in a high-repetition-rate pulse train~\cite{cundiff2010optical}.

Here, we introduce a programmable pulse-by-pulse shaper (PPS) that treats the \textit{pulse index} as an additional programmable dimension. Instead of applying a static phase mask to a repeated pulse train, the PPS assigns a distinct spectral phase function to each incoming pulse from a mode-locked fiber laser operating near 20~MHz. The system combines temporal $4f$ imaging with FPGA-synchronized electro-optic modulation, converting an electronically programmed RF waveform into a pulse-index-dependent spectral phase map. 

Using the built PPS, we demonstrate four levels of programmable pulse synthesis at the individual-pulse level. First, zero-order constant phase coding switches and scans the signal pulse, which gives pulse-by-pulse spectral-temporal double-slit like interferences by addressing one reference pulse. Second, programmable first-order phase converts ns level sliding into deterministic fs resolution movement on trajectories. Third, fractional-order GDD phase (see \cite{laskin2000fractional,malomed2021optical,liu2023experimental} for more details about fractional GDD) programming synthesizes spectral-temporal breathers, in which pulse-width breathing in the temporal domain is converted into spectral breathing in the frequency domain after one nonlinear fiber stage. Finally, by scanning the maximum fractional GDD, we construct a phase-level-dependent map of nonlinear spectral breathing, revealing transitions between weak single-envelope breathing, multi-peak spectral splitting, and strongly breathing merged-spectrum dynamics.

These results represent an important step toward the long-standing vision of on-demand \textit {pulse-by-pulse} optical arbitrary waveform generation, and establish the {pulse index} as a programmable degree of freedom for ultrafast pulse shaping and pulse-resolved nonlinear optics. In conventional nonlinear optical systems, pulse-by-pulse dynamics, such as breathing, often arise from self-organized processes governed by internal parameters, such as gain, loss, dispersion, and nonlinearity in one mode-locked fiber laser system \cite{herink2017real,kurtz2020resonant,Liu2022SM,peng2019breathing,cui2023dichromatic}. 
Although these approaches have revealed rich nonlinear dynamics, the resulting states are typically governed by the intrinsic evolution of the mode-locked fiber laser system, which are difficult to conceave on-demand synthesis at the level of individual pulses. In contrast, the PPS externally writes a programmable phase sequence onto the pulse train and maps the pulse sequence into nonlinear optical evolution. This provides a complementary route to the \textit {pulse resolved} nonlinear dynamics within one deterministically synthesized pulse by pulse manner.

\newpage
\begin{figure}[!h]
  \centering
       \includegraphics[width=18cm]{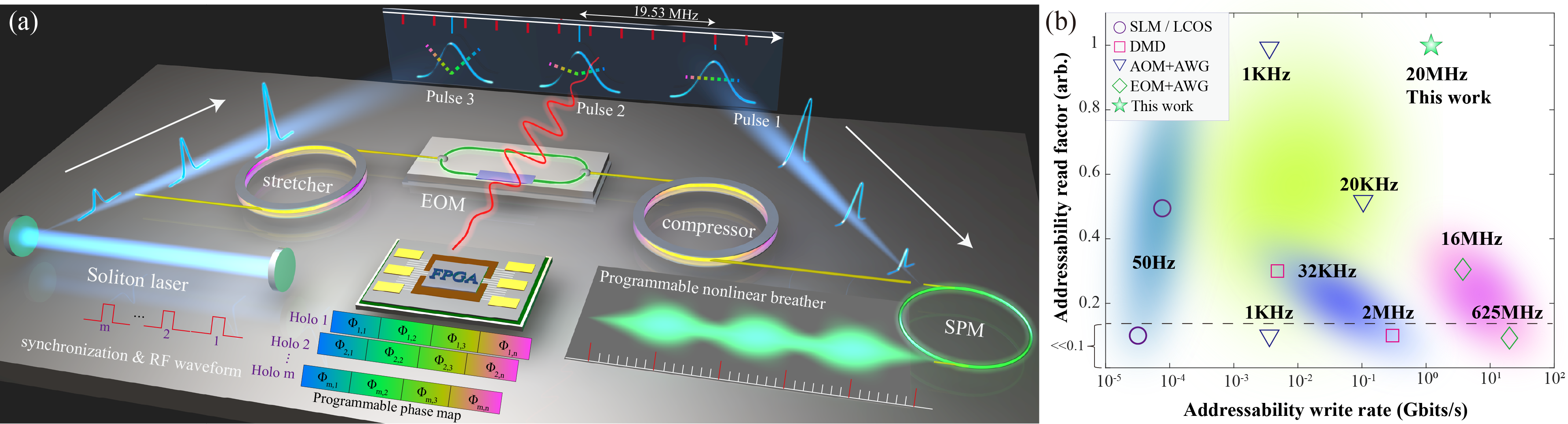}
\caption{Programmable pulse-by-pulse spectral-temporal shaper.
(a) Conceptual schematic of the pulse-by-pulse shaper. FPGA-synchronized RF waveforms drive an EOM embedded in a temporal $4f$ imaging system, enabling programmable spectral phase functions to be assigned to each of individual pulse. A subsequent nonlinear fiber stage maps the programmed pulse sequence into pulse-resolved nonlinear dynamics.
(b) Technology landscape of representative pulse-shaping platforms. The horizontal axis represents the programmable write addressability rate, defined as the product of the update rate and the control information per frame, expressed in Gbits/s.  The vertical axis represents the read addressability factor relative with write one, from quasi-static ($\ll$0.1) to pulse-by-pulse readout (=1). Colored regions indicate typical operating domains of SLM/LCOS-, DMD-, AOM/AWG-, and EOM/AWG-based pulse shapers. The star highlights the present work, which occupies the regime in both write and read MHz-rate pulse-level addressability.}
\label{F1}
\end{figure}

\section*{1. Concept for programmable pulse-by-pulse shaper}

The central concept of the programmable pulse-by-pulse shaper (PPS) is to assign a distinct spectral phase function to each pulse emitted from the mode locked fiber laser. Conventional pulse-shaping platforms generally apply a static or slowly mask to many nominally identical pulses, such that the shaped pulse train remains nearly invariant over successive round trips. In contrast, the PPS treats the \textit{pulse index n} as an additional programmable dimension. Each incoming pulse can therefore show a different profile, allowing the optical field to evolve as a designed pulse sequence rather than as a repetition of the same waveform.

Figure~\ref{F1}(a) illustrates the operating principle of the PPS based on a temporal $4f$ imaging system. A mode-locked pulse train with picosecond-scale pulse duration is first temporally stretched by the CFBG with large GDD amount, which ensures its spectral components are mapped into a nanosecond-scale time window accessible to a fast electro-optic modulation. A pulse-synchronized RF waveform generated by FPGA-based electronics drives the EOM and imprints a programmed temporal modulation onto each stretched pulse. After recompression, this temporal modulation is converted back into an effective spectral phase applied to the incident ultrashort pulse. By updating the RF waveform synchronously with the pulse index, the system implements a pulse-index-dependent spectral transfer function,
\begin{equation}
\label{E-shaper}
\tilde{E}_{\rm out}(\Omega,n)
=
\tilde{E}_{\rm in}(\Omega,n)
R(\Omega)
\exp[i\phi_{\rm EOM}(\Omega,n)],
\end{equation}
where $n$ is the pulse index and $\Omega=\omega-\omega_0$ is the frequency detuning from the carrier frequency. Here, \(R(\Omega)\) represents the spectral transfer function of the temporal imaging system, including the finite bandwidth and residual filtering response of the CFBG-based stretcher--compressor pair, and \(\phi_{\rm EOM}(\Omega,n)\) denotes the programmable phase loaded onto the \(n\)-th incoming pulse.

The programmed phase can be expanded as
\begin{equation}
\label{E-phi}
\phi_{\rm EOM}(\Omega,n)
=
\phi_{0,n}
+
\tau_{1,n}\Omega
+
\tau_{2,n}\Omega^2
+\cdots ,
\end{equation}
where each of orders correspond to different levels of pulse-by-pulse control. The zero-order term \(\phi_{0,n}\) controls the pulse-index-dependent constant phase, 
enabling phase-coded spectral interference by inducing one reference pulse. The first-order term \(\tau_{1,n}\Omega\) controls the group delay and therefore the temporal position of each pulse. Second- and higher-order terms reshape the internal temporal waveform such as pulse duration. In addition to integer-order phase terms, fractional-order spectral phases can also be programmed \cite{agrawal2000nonlinear,liu2023experimental}, providing an extended route to generate richer temporal waveform structures such as programmable temporal breathers.

The PPS therefore provides a unified framework for spectral-temporal pulse control. In the linear regime, pulse-index-dependent phase programming directly controls the phase, delay, and temporal waveform of individual pulses. When the shaped pulse train is launched into a nonlinear fiber stage, such as one dominated by self-phase modulation (SPM), the programmed temporal waveform is further converted into pulse-index-dependent spectral reshaping by inducing a SPM phase in the time domain:
\begin{equation}\label{E-SPM-phase}
E_{\rm out}(t,n)=E_{\rm in}(t,n)\exp(i\phi_{\rm SPM}),
\end{equation}
with $\phi_{\rm SPM}(t,n)=BI(t,n)$ is the temporal phase induced by SPM, in which $B(=\gamma P_{k} \delta z)$ is the nonlinear strength\cite{agrawal2000nonlinear}. As a result, variations in the temporal profile of each pulse $I(t,n)$ , such as pulse duration, can be mapped into the spectral domain on a nonlinear fiber stage. This forms the basis for programmable synthesis of spectral-temporal nonlinear pulse dynamics.

To characterize these pulse-resolved dynamics, we further implement a pulse-by-pulse reconstruction method by introducing a reference pulse in a Mach--Zehnder interferometer. Using Fourier-transform spectral interferometry~\cite{walmsley2009characterization}, the temporal profile of each addressed pulse can be reconstructed from the measured spectral interferogram. This measurement strategy allows the programmed pulse sequence to be evaluated at the individual-pulse level, rather than only through averaged spectral or temporal observables.


The positioning of this approach is summarized in Fig.~\ref{F1}(b), which compares representative pulse-shaping technologies in terms of read and write addressability rates. To quantify the write-side programmability, we define a control information rate
\begin{equation}
\label{eq:Rctrl}
R_{\rm c}
=
f_{\rm update} C_{\rm frame}
=
f_{\rm update}
N_{\rm DOF}
\log_2\left(N_{\rm level}\right).
\end{equation}
Here, \(f_{\rm update}\) denotes the rate at which independent control patterns can be applied to the optical pulse train, \(N_{\rm DOF}\) is the number of independently controllable degrees of freedom, and \(N_{\rm level}\) is the number of resolvable control levels per degree of freedom. Together, these quantities define \(C_{\rm frame}\), the accessible control information per update, expressed in bits.

For conventional frame-based pulse shapers, \(R_{\rm c}\) is typically limited by the device refresh rate. For pulse-by-pulse shaping systems, the effective update rate can approach the repetition rate of the pulse train, provided that individual pulses can be independently addressed. Spatial-light-modulator and LCOS-based shapers provide high control dimensionality but are generally limited by slow refresh rates $\sim$ 50Hz. DMD-, AOM-, and AWG-assisted platforms can increase the writing speed in tens of KHz to MHz, but their pulse-level write and read addressability is often restricted to static, burst-level, or partially addressed operation. The present PPS combines MHz-rate pulse-level read addressability with high-speed electronic writing, enabling programmable optical waveforms to be updated at the individual-pulse level. This capability places the system in a regime suitable for pulse-by-pulse spectral-temporal control and programmable nonlinear breathing dynamics.

\newpage
\begin{figure}[!h]
  \centering
       \includegraphics[width=17cm]{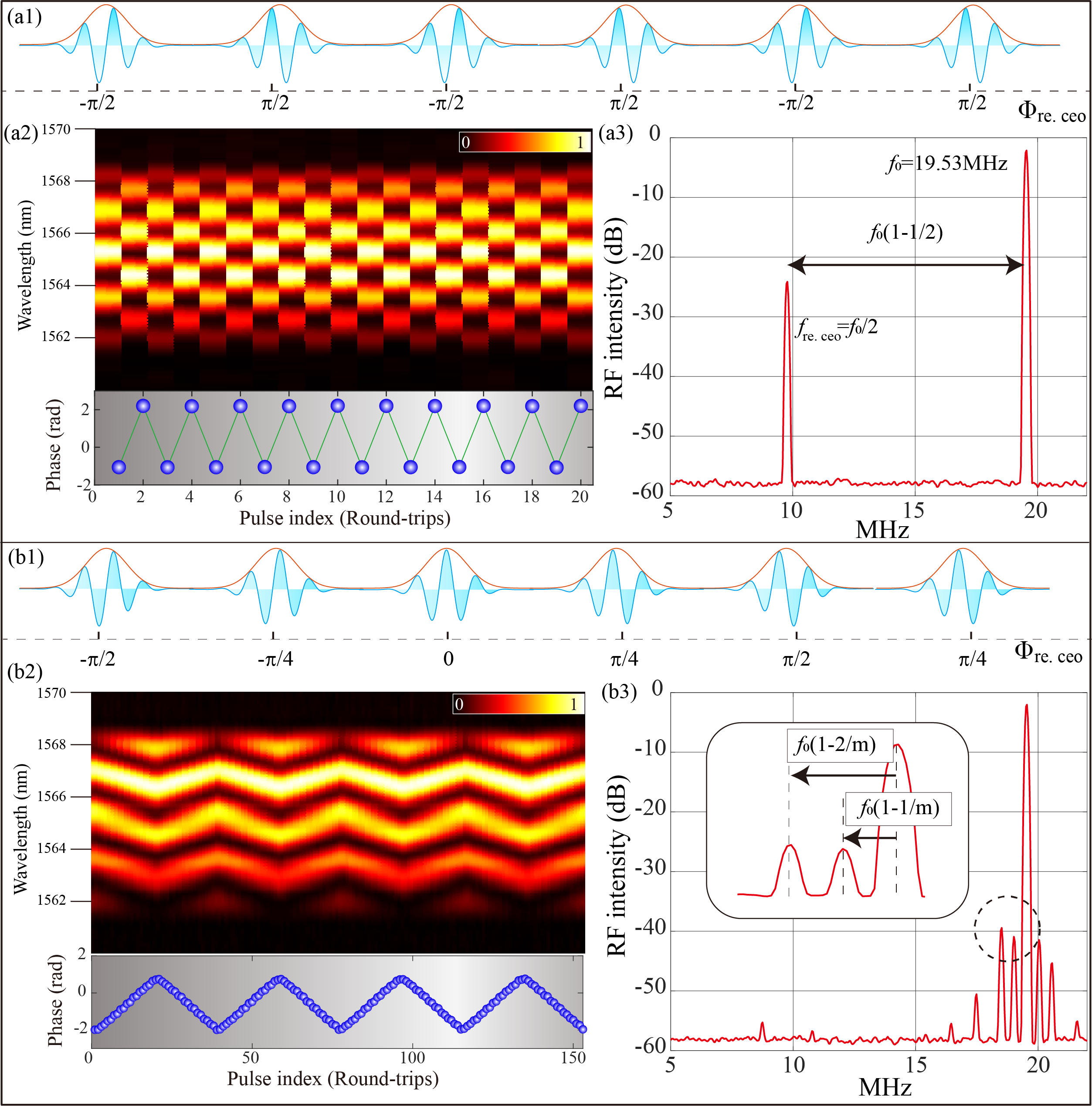}
\caption{
Pulse-by-pulse zero-order optical phase programming.
(a) Binary $\{-\pi/2 \rightarrow \pi/2 \rightarrow  -\pi/2 \rightarrow \pi/2 \}$ phase switching.
(a1) Carrier-field illustration of adjacent pulses with opposite zero-order optical phase. The intensity envelope is unchanged, whereas the electric field is reversed.
(a2) Measured single-shot spectral interferogram and reconstructed pulse-index-resolved phase for the binary phase sequence.
(a3) RF spectrum showing the pulse repetition frequency \(f_0=19.53~\mathrm{MHz}\) and the subharmonic component at \(f_0/2=9.765~\mathrm{MHz}\), corresponding to the phase modulation period.
(b) Multi-level periodic zero-order phase programming $\{-\pi/2   \cdot  \cdot  \cdot \rightarrow 0  \cdot  \cdot  \cdot \rightarrow \pi/2\}$.
(b1) Carrier-field illustration of a periodically stepped optical phase.
(b2) Measured single-shot spectral interferogram and reconstructed phase trajectory.
(b3) RF spectrum showing subharmonic components determined by the programmed phase period. In the few-cycle limit, this zero-order optical phase corresponds to the carrier phase of the signal pulse relative to the reference pulse, providing a route toward pulse-by-pulse relative CEP control.
}\label{F2}
\end{figure}

\section*{2. Pulse-by-pulse double-slit like interference in the temporal-spectral domain}
We first demonstrate pulse-by-pulse addressability by programming the zero-order phase of a temporally separated two-pulse waveform. In this configuration, one of pulses  is the signal modulated by EOM and the other is the reference without modulation. Two pulse replicas act as a temporal analogue of a double slit \cite{tirole2023double}: their relative delay determines the spectral-fringe period, whereas the relative zero-order phase determines the fringe offset. Therefore, $n$-th pulse-resolved spectral intensity can be written as
\begin{equation}
I_n(\Omega,n)\sim 1+\cos{(\Omega\tau+\phi_{0,n})}
\end{equation}
where $\tau$ is the temporal separation between the two pulse replicas and $\phi_{0,n}$ is the pulse-index-dependent zero-order phase. Therefore, pulse to pulse changing $\phi_{0,n}$ directly shifts the spectral interference fringes without changing the pulse separation.

Figure~\ref{F2}(a) shows binary zero-order phase coding, that is $\{-\pi/2 \rightarrow \pi/2 \rightarrow -\pi/2 \rightarrow \pi/2 \cdot \cdot \cdot \}$. The programmed phase alternates between two values separated by $\pi$, corresponding to a pulse-by-pulse reversal of the temporal double-slit like interference. As illustrated in Fig.~\ref{F2}(a1), this operation is equivalent to assigning alternating carrier-envelope phases(CEP) to successive pulses \cite{jones2000carrier}. The measured pulse-resolved spectra in Fig.~\ref{F2}(a2) show that the interference fringes switch between two complementary states from one pulse to the next, in agreement with the programmed $0/\pi$ phase sequence. This periodic phase reversal produces a clear subharmonic component in the RF spectrum at $f_{\rm re,ceo}=f_0/2$, where $f_0=19.53$~MHz is the laser repetition rate [Fig.~\ref{F2}(a3)]. The appearance of this RF component confirms that the optical interference state is switched at half of the repetition frequency.

We further extend the zero-order phase coding from binary switching to multilevel periodic constant phase modulation. As shown in Fig.~\ref{F2}(b), the programmed phase is scanned linearly from $-\pi/2$ to $\pi/2$ and repeat over a period of $m$ pulses, forming a triangular pulse-index-dependent phase trajectory [Fig.~\ref{F2}(b1)]. The measured pulse-resolved spectra exhibit a corresponding \textit{zigzag} motion of the interference fringes [Fig.~\ref{F2}(b2)], showing that the spectral phase offset can be continuously addressed over successive pulses. In the RF domain, this periodic phase modulation gives rise to a set of sidebands around the repetition-frequency component [Fig.~\ref{F2}(b3)]. The sideband spacing is determined by the phase-coding period and follows $f_0/m$, with characteristic components appearing near $f_0(1-1/m)$, $f_0(1-2/m)$, etc.. The periodic \(m\)-pulse phase sequence produces RF subharmonic components determined by the programmed phase period, which provides an RF-domain signature of the pulse-by-pulse zero-order phase programming and suggests an active locking route toward relative CEP control for pulse trains.

The temporal-spectral correspondence establishes a direct analogy for the double-slit interference patterns. These patterns could be changed at the level of repetitive frequency by employing the programmed pulse-index-dependent zero-order phase, and thus produce the subharmonic in RF-domain. Similar spectral-interference patterns were also observed in double pulse regime, such as soliton molecular experiments, in which the strategies includes adjustments of the dispersion\cite{Liu2022SM}, time varying gain\cite{herink2017real,kurtz2020resonant}, as well as polarization in the cavity \cite{krupa2017real,liu2018real}.

\section*{3. Pulse-by-pulse temporal trajectory synthesis by first-order phase programming}

After demonstrating zero-order phase coding of temporal double-slit spectral interference, we next use the PPS to program the first-order spectral phase of individual pulses. A first-order phase term provides direct control over the group delay of each pulse. For the $n$-th pulse, the programmed phase and the induced temporal delay can be written as
\begin{equation}
\label{E-1st}
\begin{split}
\phi_1(\Omega,n) &= \tau(n)\Omega,\\
\tau(n) &= \frac{\partial\phi_1(\Omega,n)}{\partial\Omega},
\end{split}
\end{equation}
where $\Omega=\omega-\omega_0$ is the angular-frequency detuning from the carrier frequency, and $\tau(n)$ is the pulse-index-dependent group delay. Therefore, by assigning a different first-order phase slope to each pulse, the PPS translates an electronically programmed phase sequence into a deterministic temporal moving trajectory of the optical pulse train.

\begin{figure}[!h]
  \centering
       \includegraphics[width=17cm]{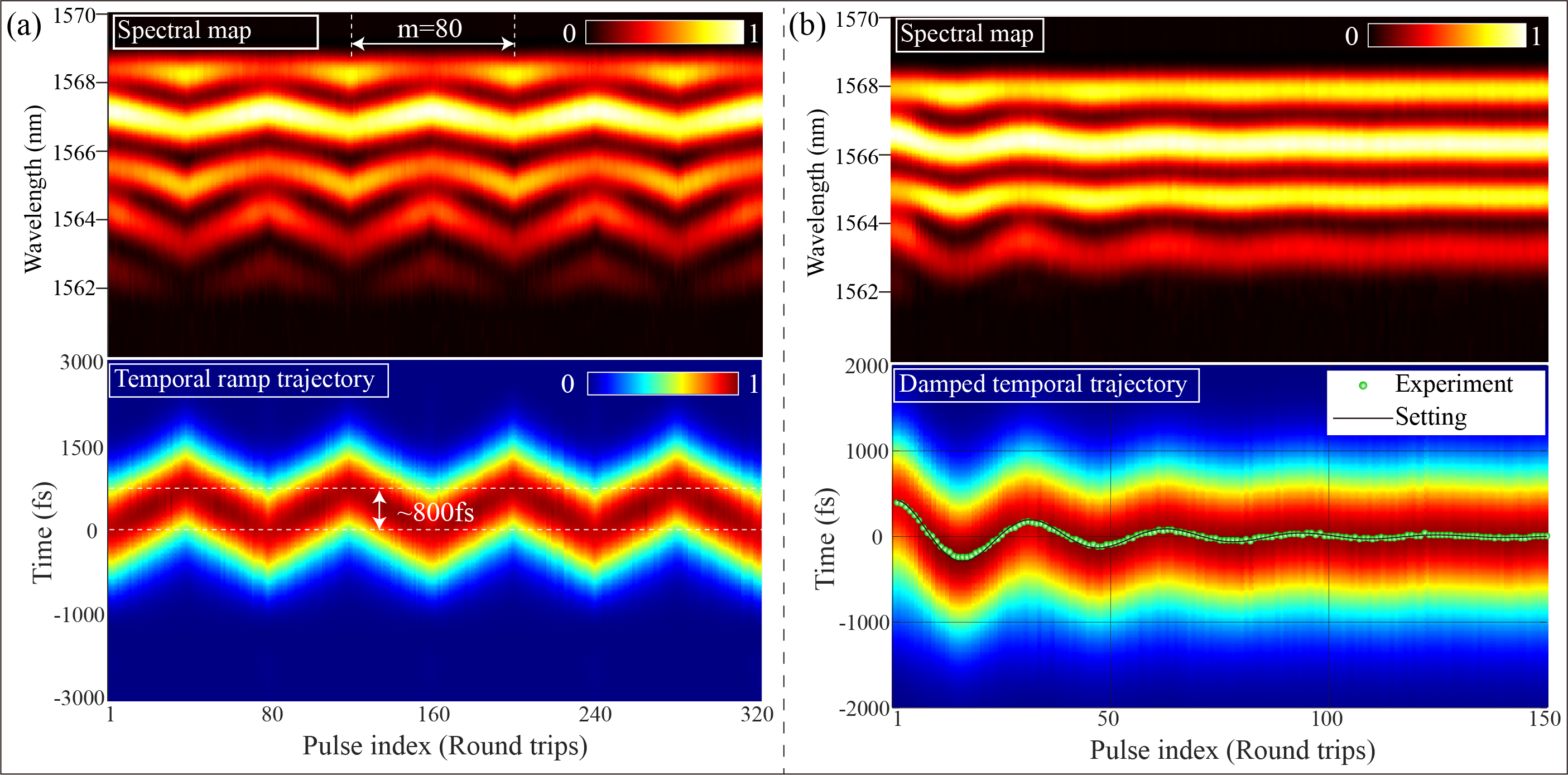}
\caption{
Programmable temporal trajectories enabled by pulse-by-pulse first-order phase control.
A first-order spectral phase \(\phi_1(\Omega,n)=\tau(n)\Omega\) shifts the \(n\)-th pulse in time by \(\tau(n)=\partial \phi_1/\partial \Omega\).
(a) Periodic triangular temporal trajectory. The measured spectral map  (top panel) and the reconstructed temporal intensity map (bottom panel); it reveals a periodic temporal ramp with \(m=80\) pulses per cycle and an excursion of approximately \(800~\mathrm{fs}\).
(b) Damped oscillatory temporal trajectory. The measured spectral map (top panel) and retrieved temporal trajectory (bottom panel). Green markers represent the experimentally retrieved delay, while the black curve shows the programmed setting.
}
\label{F3}
\end{figure}

Figure~\ref{F3} demonstrates two representative pulse-by-pulse temporal trajectories. In Fig.~\ref{F3}(a), a periodic triangular delay sequence is programmed with a period of $m=80$ pulses. The measured pulse-resolved spectral map shows a periodic shift of the spectral features, while the reconstructed temporal intensity map reveals a corresponding triangular temporal motion. The extracted trajectory exhibits a temporal excursion of approximately 800~fs, confirming that the first-order phase slope is mapped into a pulse-index-dependent delay. This measurement shows that the PPS can generate a periodic temporal ramp trajectory over successive pulses without mechanically changing the optical path.

The system can also operate and record over a much longer pulse sequence. Under the same first-order phase encoding principle, we increased the step resolution and programmed a slowly varying linear phase sequence over $m=5000$ pulses on EOM. This sequence yielded a continuous temporal displacement of approximately 855~fs. The corresponding programmable fitted delay resolution is about 170~as, which is lower than the retrieved timing jitter $\sim$ 38~fs of the used homemade fiber laser. More details on the long-sequence calibration and timing analysis are provided in the \textbf{Supplementary Information}.

To further demonstrate arbitrary temporal-trajectory synthesis, the target delay movement is set to a damped oscillatory function,
\begin{equation}
\label{E-Damp}
\tau_{\mathrm{tar}}(n)
=
\tau_{\mathrm{tar},0}
\cos\left[C_1\theta_{\mathrm{tar}}(n)\right]
\exp\left[-\frac{\theta_{\mathrm{tar}}(n)}{C_2}\right],
\end{equation}
where $\theta_{\mathrm{tar}}(n)$ is free parameter that scans from 0 to $2\pi$, and $C_1$ and $C_2$ determine the oscillation frequency and damping strength, respectively. The target delay $\tau_{\mathrm{tar}}(n)$ is implemented through the first-order phase relation defined in Eq.~\ref{E-1st}, with the maximum delay amplitude $\tau_{\mathrm{tar},0}$ set to 427~fs.

Figure~\ref{F3}(b) shows the spectral and temporal intensity maps obtained for this damped oscillatory phase program. The reconstructed temporal trajectory follows the programmed setting, as indicated by the agreement between the experimentally retrieved delay points, shown by green markers, and the target curve, shown by the black line. This result shows that the PPS is not limited to simple periodic ramps, but can synthesize user-defined temporal trajectories on a pulse-by-pulse basis.


\begin{figure}[!h]
  \centering
       \includegraphics[width=17.5cm]{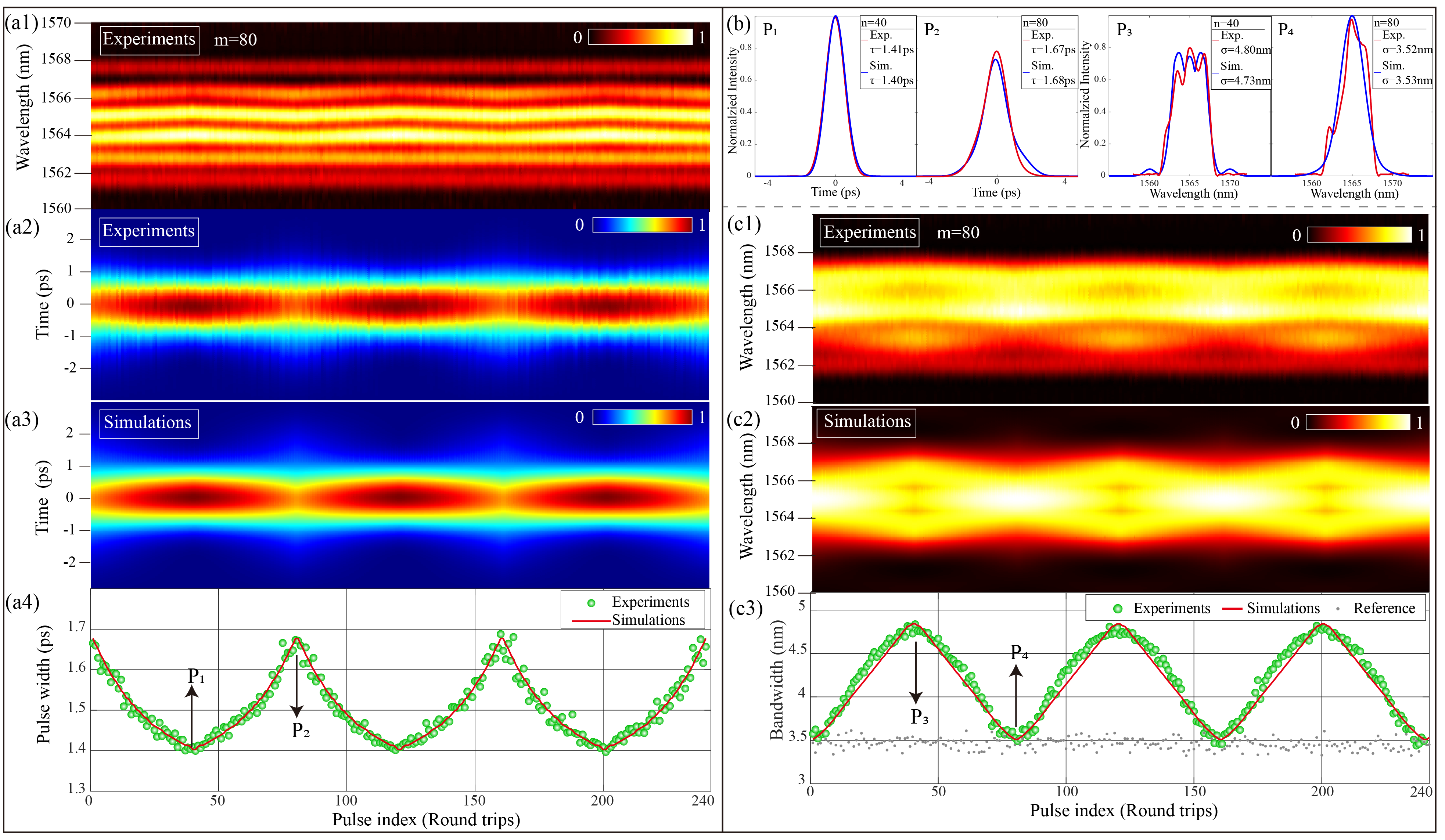}
\caption{Programmable spectral-temporal breathing after a nonlinear fiber stage.
(a) Fractional-order phase programming generates a \textit{temporal breather} with a period of $m=80$ pulses. The measured pulse-resolved spectral intensity (a1), reconstructed temporal intensity in experiment (a2) and simulation (a3), and extracted pulse width (a4) show periodic temporal breathing.
(b) Representative temporal and spectral breathing states. $P_1$ and $P_2$ correspond to the minimum and maximum pulse widths in (a4), whereas $P_3$ and $P_4$ correspond to spectral breathing states with the maximum and minimum bandwidth (c3).
(c) Nonlinear spectral breathing after fiber propagation. Measured (c1) and simulated (c2) pulse-resolved spectra are quantified by the extracted spectral bandwidth in (c3), with the unmodulated reference shown in gray. }
\label{F4}
\end{figure}

\section*{4. Pulse-by-pulse programmable synthesis of spectral-temporal breathers}

We next extend the pulse-by-pulse control from temporal displacement to temporal breathing dynamics by programming a fractional-order spectral phase with L\'{e}vy index (LI)=1 \cite{laskin2000fractionalpath}. Compared with arbitrary high-resolution waveform synthesis like second- or high- order spectral phase, this operation can be implemented efficiently with the available AWG bandwidth, since the desired modulation is described by a low-dimensional fractional-phase function, such as three points to define the phase profile for LI=1, rather than by a phase with arbitrary curves that needs several tens of points to fit \cite{malomed2021optical,hoang2026nyquist}. In contrast to the first-order spectral phase, which maps directly onto a group delay and therefore shifts the pulse position in time, a fractional-order spectral phase modifies the internal structure of each pulse in the temporal domain. The corresponding pulse-index-dependent spectral phase engineering process can be written as
\begin{equation}
\label{E-FGVD}
\tilde{E}_{\mathrm{out}}(\Omega,n)
=\tilde{E}_{\mathrm{in}}(\Omega,n)
\sqrt{\delta}\,
\mathrm{Rect}\left({\Omega}/{\Omega_m}\right)
\exp\left[i C_F(n)|\Omega|^{\alpha}\right],
\end{equation}
where $\delta$ accounts for the insertion loss, and $\mathrm{Rect}(\Omega/\Omega_m)$ represents the finite spectral window of the shaping system. The coefficient $C_F(n)$ determines the pulse-index-dependent fractional-phase strength, with the maximum programmed phase denoted by $\Phi_m$. The exponent $\alpha$, also referred to as the L\'{e}vy index, determines the order of the fractional phase \cite{laskin2000fractionalpath}. In particular, the case $\alpha=1$ introduces a non-quadratic spectral phase that can first stretch the pulse duration and then form a double-lobe temporal waveform as the increase of $C_F(n)$ ~\cite{liu2023experimental,hoang2025nonlinear}.  By periodically varying the programmed fractional-phase strength over successive pulses, the PPS generates a pulse-index-dependent \textit{temporal breathing} trajectory.

Figure~\ref{F4} shows the spectral-temporal breathing dynamics at a representative programmed phase condition, with $\alpha=1$ and $\Phi_m=\pi/3$. The fractional-order phase sequence is applied with a breathing period of $m=80$ pulses. Before the nonlinear fiber stage, the pulse-resolved spectral intensity exhibits a periodic modulation over the breathing cycle [Fig.~\ref{F4}(a1)]. The reconstructed temporal intensity maps from experiment and simulation reveal a periodic variation of the pulse duration [Figs.~\ref{F4}(a2,a3)], confirming the formation of a programmable temporal breather. The extracted pulse width agrees well with the simulation [Fig.~\ref{F4}(a4)]. Two marked positions $P_1$ and $P_2$ correspond to the maximum and minimum temporal-width states within one breathing cycle, with pulse widths of 1.41 ps and 1.67 ps, respectively. The representative temporal profiles in Fig.~\ref{F4}(b) further show the corresponding pulse-envelope evolution at these two states, located at pulse indices $n=40$ and $n=80$.

When the shaped pulse train is subsequently launched into the nonlinear fiber stage, the temporal breather acts as a pulse-index-dependent input to the nonlinear propagation process. The programmed temporal breathing is then converted into a corresponding \textit{spectral breathing} response through nonlinear effects, dominated here by self-phase modulation (SPM). This process can be understood as a pulse-resolved single-pulse propagation problem, in which each pulse in the sequence carries a distinct temporal waveform $\Psi(t,n)$. The propagation in the gain and passive fiber sections can be described by a pulse resolved nonlinear Schr\"{o}dinger-type model,
\begin{equation}
\label{E-SPM}
i\frac{\partial E(t,z,n)}{\partial z}
=
\left(
\frac{\beta_2}{2}\frac{\partial^2}{\partial t^2}
+
i\frac{g}{2}
\right)
E(t,z,n)
-
\gamma |E(t,z,n)|^2E(t,z,n),
\end{equation}
where $\beta_2$ is the second-order dispersion coefficient, $\gamma$ is the nonlinear coefficient, and $g$ represents the optical gain. For the erbium-doped fiber section, the gain can be modeled as $g=g_0\exp\left[-E(z)/E_{\mathrm{sat}}\right]$, where $g_0$ is the small-signal gain, $E_{\mathrm{sat}}$ is the saturation energy, and $E(z)=\int_{-\infty}^{+\infty}|\Psi(t,z,n)|^2 dt$
is the pulse energy. In the present regime, the EDFA mainly sets the pulse energy launched into the following nonlinear fiber, while the dominant spectral reshaping arises from SPM. The accumulated nonlinear phase for the $n$-th pulse can therefore be estimated as $\Phi_{\mathrm{NL},n}\simeq \gamma P_{0,n}L_{\mathrm{eff}}$,
where $P_{0,n}$ is the peak power of the $n$-th shaped pulse and $L_{\mathrm{eff}}$ is the effective nonlinear length. Since the programmed fractional phase changes the pulse duration via pulse to pulse, both $P_{0,n}$ and the instantaneous nonlinear frequency shift become pulse-index dependent. This allows the programmed temporal breathing to map into a nonlinear spectral-breathing response.

The measured and simulated pulse-resolved spectra after the nonlinear fiber stage show a periodic expansion and contraction of the output spectrum [Figs.~\ref{F4}(c1,c2)]. This nonlinear spectral breathing is quantified by the extracted spectral bandwidth in Fig.~\ref{F4}(c3). Both experiments and simulations show a clear periodic bandwidth modulation, whereas the unmodulated reference remains nearly constant, as indicated by the gray points. Representative spectra corresponding to bandwidth with the maximum and minimum are shown in Fig.~\ref{F4}(b) as $P_3$ and $P_4$, respectively. These results establish a direct link between the programmed fractional-order phase, the resulting temporal breathing, and the nonlinear spectral response. The representative phase level used in Fig.~\ref{F4} corresponds to the horizontal dashed line in Fig.~\ref{F5}(a1,b1), where the nonlinear breathing dynamics are further mapped over a broader phase-level space.

\section*{5. Phase-level-dependent diagram of nonlinear spectral breathing}

\begin{figure}[!h]
  \centering
  \includegraphics[width=17.5cm]{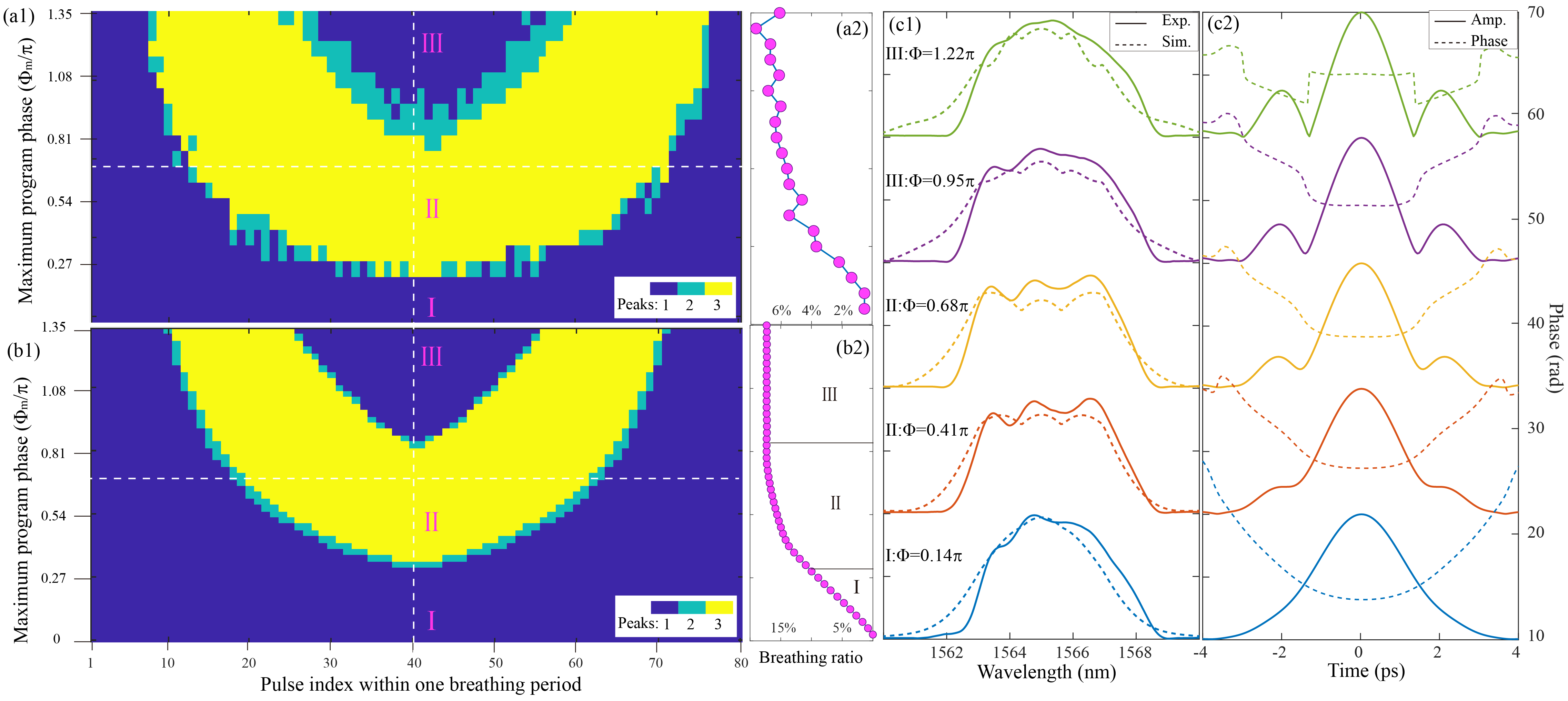}
\caption{Phase-level-dependent morphology diagram of programmable nonlinear spectral breathing.
(a,b) Experimental (a1) and simulated (b1) morphology maps over one 80-pulse breathing period, obtained by varying the maximum programmed fractional-phase level $\Phi_m/\pi$. The colors denote three spectral-morphology regimes: I, weakly modulated single-envelope breathing; II, multi-peak spectral splitting; and III, strongly breathing merged-spectrum dynamics. The horizontal dashed line indicates the representative phase level used in Fig.~\ref{F4}, and the vertical dashed line marks the pulse index $N=40$ used in (c).
(a2,b2) Extracted breathing ratio from experiment and simulation, respectively.
(c1) Representative output spectra at $N=40$ for selected phase levels, with solid and dashed curves denoting experiment and simulation.
(c2) Simulated temporal amplitude and phase profiles corresponding to the spectra in (c1), revealing the temporal origin of the observed spectral breather morphology transitions.}
  \label{F5}
\end{figure}

The programmable nature of the PPS enables the nonlinear spectral breather to be explored beyond a single operating point. We therefore varied the maximum programmed fractional-phase level, $\Phi_m$, and recorded the pulse-resolved output spectra over one 80-pulse breathing period. For each pulse index and programmed phase level, the output spectrum was classified according to its spectral morphology, using the resolved peak number together with the corresponding breathing ratio. This yields a phase-level-dependent regime map of the nonlinear spectral breather, as shown in Fig.~\ref{F5}.

In the experiment, a relatively narrow spectral filtering window of approximately 7 nm was used in the temporal $4f$ shaper to suppress weak non-dominant side peaks, such as those observed in the broad-bandwidth state $P_4$ in Fig.~\ref{F4}(b). This choice trades part of the available shaping bandwidth for cleaner spectral profiles and more robust morphology classification. The experimental regime map in Fig.~\ref{F5}(a1) reveals three distinct morphology regimes around the center of the breathing cycle, near pulse index $n=40$. The classification is based on the number of resolved spectral peaks and the breathing strength, with a minimum peak prominence of 2.5\% used to avoid counting weak noise-induced or residual side peaks.

At low programmed phase levels, $\Phi_m \in [0,0.3\pi]$, the output remains in regime I, corresponding to a weakly modulated single-envelope regime with limited spectral reshaping and small breathing contrast. As $\Phi_m$ increases toward approximately $0.8\pi$, a broad regime II emerges, where the nonlinear spectra exhibit multi-peak spectral splitting over a large portion of the breathing period. At still larger phase levels, up to $\Phi_m=1.35\pi$, the central region of the breathing cycle enters regime III. This regime should not be interpreted as a higher peak-number state. Instead, it corresponds to a strongly breathing merged-spectrum regime, where the previously resolved multi-peak structures merge into a dominant spectral envelope while the breathing ratio remains large. Thus, regimes I and III can both exhibit a single dominant spectral envelope, but they represent physically distinct states: weakly modulated breathing in regime I and strongly nonlinear spectral reshaping in regime III.

Based on the fractional-phase filtering model in Eq.~\ref{E-FGVD} and the nonlinear propagation model in Eq.~\ref{E-SPM}, we performed simulations using the same F-GVD phase settings as in the experiment. The simulated regime map in Fig.~\ref{F5}(b1) reproduces the main features of the experimental diagram, including the expansion of the multi-peak splitting regime and the emergence of the strongly breathing merged-spectrum regime at higher phase levels. The smoother boundaries in the simulation and the broadened transitions in the experiment are consistent with finite spectral resolution, residual system ripples, measurement noise, and the threshold-based peak-classification procedure.

To quantify the continuous component of the nonlinear breathing, we extract the breathing ratio,
\[
R_{\rm B}
=
\frac{\Delta\lambda_{\max}-\Delta\lambda_{\min}}
{\Delta\lambda_{\max}+\Delta\lambda_{\min}},
\]
where $\Delta\lambda$ denotes the FWHM spectral bandwidth extracted over one 80-pulse breathing period. The breathing-ratio traces in Figs.~\ref{F5}(a2,b2) show that the spectral breathing contrast increases systematically with the programmed phase level. In regime I, the breathing ratio increases approximately linearly with $\Phi_m$. It then enters a transition region associated with the multi-peak splitting regime II, where the increase becomes slower. At higher phase levels, the response approaches the strongly breathing merged-spectrum regime III, where the breathing ratio remains large while the spectral morphology returns to a dominant single-envelope structure.

Representative spectra extracted at pulse index $n=40$ provide a direct view of the underlying spectral evolution [Fig.~\ref{F5}(c1)]. As the programmed phase level $\Phi_m$ increases, the output spectrum evolves from a weak single-envelope profile to a multi-peak split spectrum and finally to a broadened merged-envelope state. The experimentally measured spectra follow the same trend as the simulations, supporting the interpretation of the regime map. The corresponding simulated temporal amplitude and phase profiles [Fig.~\ref{F5}(c2)] further reveal the origin of these transitions. Increasing $\Phi_m$, together with nonlinear propagation in the fiber stage, reshapes both the temporal intensity envelope and the temporal phase, thereby modifying the instantaneous nonlinear frequency shift. This temporal amplitude--phase evolution provides the mechanism by which pulse-by-pulse F-GVD phase programming is converted into distinct nonlinear spectral-breathing morphologies.

\section*{Discussion}

The results presented here demonstrate a programmable route to pulse-by-pulse ultrafast optical waveform synthesis. A long-standing goal in ultrafast pulse shaping and optical arbitrary waveform generation is to move beyond static or slowly refreshed masks and to assign independent optical waveforms to successive pulses in a high-repetition-rate pulse train. The PPS addresses this goal by treating the pulse index as an additional programmable dimension. In this framework, the optical field is no longer a simple repetition of identical pulses; instead, it becomes a designed sequence of individually programmed spectral-temporal waveforms. The zero-order, first-order, and fractional-order phase demonstrations show that this concept can be applied across different levels of control, from constant phase-coded spectral interference and temporal trajectory synthesis to programmable waveform breathing.

An important consequence of this capability is that nonlinear pulse shaping can also be brought into the pulse-resolved regime. In conventional nonlinear fiber or laser systems, nonlinear dynamics such as breathing, spectral splitting, and waveform reshaping are usually governed by the intrinsic evolution of the nonlinear laser system and are tuned through global parameters such as pump power, polarization, dispersion, or cavity detuning \cite{herink2017real,kurtz2020resonant,Liu2022SM,peng2019breathing,cui2023dichromatic}. In contrast, the PPS externally writes a designed phase function onto each pulse before nonlinear propagation. Each pulse therefore enters the nonlinear fiber stage with a distinct temporal waveform and subsequently undergoes a distinct pulse resolved nonlinear propagation trajectory. This converts pulse-by-pulse linear phase programming into pulse-resolved nonlinear spectral-temporal dynamics. The observed nonlinear spectral breathing and the phase-level-dependent regime map show that breathing, splitting, and merged-spectrum states can be synthesized, measured, and classified at the individual-pulse level.

The present implementation is mainly limited by the electronic sampling resolution available for writing the phase pattern onto the temporally stretched pulse. In our current system, the FPGA operates the output waveform at an effective sampling rate of 1.25~GSa/s, which is sufficient for the phase functions demonstrated here but still limits the spectral programming resolution and the complexity of arbitrary waveform synthesis. Future implementations based on faster FPGA/DAC architectures, higher-bandwidth electro-optic modulators, and improved synchronization electronics could substantially increase the number of controllable temporal samples across the stretched pulse. This would improve the spectral phase resolution, expand the accessible control dimensionality, and enable more complex pulse-by-pulse amplitude-and-phase waveform synthesis.

\end{document}